\documentclass[12pt,a4]{article}

\usepackage{setspace}
\usepackage{geometry}
\usepackage{amsthm,amsmath,latexsym,amssymb,amsfonts,amscd}
\usepackage{graphics,lscape,fancyhdr,array,stmaryrd,euscript}
\usepackage{empheq}
\usepackage{dsfont}
\usepackage{verbatim}
\usepackage{color,tikz,tikz-cd}
\usetikzlibrary[snakes]
\usepackage{relsize,slashed}
\numberwithin{equation}{section}
\usepackage{hyperref}
\usepackage{xcolor}
\usepackage{setspace}

 \usepackage[numbers,sort&compress]{natbib}
 \usepackage[nottoc]{tocbibind}
\date{}
\newcommand{\besubeqs}{\begin{subequations}}
\newcommand{\esubeqs}{\end{subequations}}

\newcommand{\mak}{\mathcal{K}}

\begin{document}

\title{ Holography of new conformal higher spin gravities in 3d for low spins}
\maketitle
\vspace{-1.9cm}
\begin{center}
\author{ Iva Lovrekovi\'c\footnote{email: lovrekovic@hep.itp.tuwien.ac.at}}\\
{\em
Institute for Theoretical Physics, TU Wien, \\
 Wiedner Hauptstr. 8-10, 1040 Vienna, Austria}\\ 
    \texttt{Contribution to Seventeenth Marcel Grossmann Meeting}
 \end{center}  

\begin{abstract}
We study holography of the 3d Chern-Simons theory as a gauge theory of $so(3,2)$, $sl_4$ and $sl_5$ algebras. For the near horizon boundary conditions we present solutions from several projectors from Chern-Simons to the metric formulation. These solutions are generalized BTZ solutions for our theories. We also study the classification according to $so(3,2)$ one parameter subgroups and classify obtained solutions.
\end{abstract}

In this talk we will introduce one of the candidates of quantum gravity, which is called higher spin gravity (HiSGRA). 
We will motivate the research in general, and present one example of the higher spin (HS) models. This example is going to be in the lower number of dimensions, in three dimensions (3d), and it is going to be based on the Chern-Simons (CS) action.
Beside key aspects such as the motivation behind studying HiSGRA, and its connections to CS theory, we will talk about the holography of HiSGRA, black hole solutions, and one possible classification of these solutions.

We know that General relativity (GR) has been experimentally verified to extreme limits. This is evident by the detection of gravitational waves (by the LIGO and Virgo collaborations) and the first image of a black hole (by the Event Horizon Telescope, 2019). 
However, there are still unresolved questions in both theoretical and observational side of gravitational physics. The main unresolved theoretical questions in Einstein gravity (EG) are that EG is two-loop non-renormalizable, and that we do not know for certain which is the right way to write it as a quantum gravity theory.  EG also does not explain the galactic rotation curves  without the addition of extra dark matter particles, nor gives us candidates for these particles, and it does not actually explain what the cosmological constant is. 

For these reasons, some of the scientists have turned towards alternative models which would resolve the theoretical question of EG, and explain the phenomena EG does not explain. 
One of these theories is HiSGRA.
HiSGRA proposes a framework that extends beyond the Standard Model and EG by incorporating an infinite tower of higher-spin particles, starting from spin-0 up to an arbitrary spin. 
The reason for adding infinitely many particles is that it was realised if we want to add only one HS particle, we have to add all of them in order for the theory to be consistent. 
The first such theory that was constructed, was the theory of massless HS fields which are non-interacting. These fields are also called Fronsdal fields. Further efforts brought to construction of conformal HS theory, and the development of the Vasiliev formulation. This is a special formulation which describes an infinite tower of interacting HS fields. The same description within the usual Lagrangian formalism is still lacking. 

The approach using HiSGRA models aims to improve the quantum behaviour of gravity by allowing consistent interactions between an infinite number of HS particles. 
One of the technical difficulties of the theory is to write the action with all possible interacting fields, because the actions become more and more complicated with adding each field of the higher spin. Another technical difficulty are the no-go theorems which need to be circumvented when trying to construct HS theory on the flat space.  
 (i)  Weinberg's low energy theorem tells us that there is a clash between the Lorentz invariance of the amplitude with n-particles and one spin-s particle softly attached, and the conservation of momenta, unless all the HS particles couple universally to gravity. (ii) Coleman-Mandula theorem restricts the possible symmetries in a relativistic quantum field theory with a non-trivial S-matrix. It shows that the only allowed symmetries are the direct product of internal symmetries and the Poincar\'e group.
Since higher-spin symmetries mix spacetime and internal symmetries, the Coleman-Mandula theorem implies that such symmetries are incompatible with standard particle interactions in a conventional quantum field theory framework in flat spacetime.
(iii) The no-go theorem by Aragone-Deser does not allow for minimal gravitational interactions of massless particles of higher spin. This is due to manifest Lorentz covariance and gauge invariant field description using Fronsdal fields. 

Since 2017, the rapid development of new HiSGRA models has started. It is based on the recognition of the symmetries which allow for the construction of the Lagrangian, in some cases truncation of the infinite tower of the HS fields, and simultaneous circumvention of the no-go theorems. 
Here, the first discovered theory was Chiral HiSGRA formulated in the light-front formulation. It overcomes the Weinberg's no-go theorem since the vertices do not have a manifestly Lorentz covariant form, while in the proof of the theorem we have to use explicitly Lorentz covariant vertices. The Coleman-Mandula theorem was avoided by assuming that symmetry generators transform as spinors. The Aragone-Deser theorem is also avoided due to light-cone approach. The argument in the derivation of the theorem uses manifestly Lorentz covariant methods, which are avoided in light-cone approach.

Further theories that were discovered since that time, are new conformal HiSGRA, and partially massless HiSGRA, both in 3d. In 3d we can use the topological Chern-Simons action and we do not have propagating degrees of freedom. This construction is also such that it circumvents the no-go theorems. 
 There were further models such as massive HiSGRA in 3d and models that are derived from these models. The HiSGRA example on which we are going to focus here, is new conformal HiSGRA in 3d.  One distinctive thing about the new conformal HiSGRA is that among the theories with infinite number of HS fields, this is a construction with an infinite tower of theories, each with finite numbers of HS fields.

\section{New conformal HiSGRA in 3d}

The construction used for the 3D conformal HiSGRA is based on Chern-Simons (CS) theory \cite{Chern:1974ft,Dunne:1998qy}. CS theory serves as a foundation for numerous studies in different research areas. A special property of the theory is that it is topological, allowing us to write a gauge theory for any group 
$G$ that we wish to consider.

EG in 3d with cosmological constant can be written in  Chern-Simons form, as a gauge theory of the $so(2,2)$ algebra. Without the cosmological constant, the gauge group of EG in 3d is $ISO(2,1)$ \cite{Witten:1988hc}. The theory is represented with an action 
\begin{align}
S(\omega)=\int Tr\left( \omega\wedge d\omega+\frac{2}{3}\omega\wedge\omega\wedge\omega\right)
\end{align}
in which $\omega$ is a Lie-algebra valued 1-form. For the new conformal HiSGRA \cite{Grigoriev:2019xmp}, the algebra used to evaluate $\omega$ comes from one of the algebras in the infinite tower of algebras in the following construction:
The construction evaluates $End(V)=V\otimes V^*$, where $V$ is a vector decomposed in the modules of $so(3,2)$ algebra. 
The gauge parameter of the theory is a Lie algebra-valued 0-form, and it is valued in the same algebra.

By varying the action we obtain the equation of motion for $\omega$ decomposed for each of the generators. These are expressions for the curvatures. The expressions for the gauge transformations we decompose in the analogous way, for each of the generators. This, together with equations of motion, lets us calculate the dynamical field that remains in the theory after the gauge has been fixed. For the spin-2 theory in this framework the remaining field is conformal graviton, and the theory corresponds to conformal gravity theory in 3D.

If the theory is a gauge theory of one of the other algebras in the tower of HS algebras in our construction, we will get another field in addition to conformal graviton. For example, if we have a gauge theory of $sl(4)$ group, decomposed to distinguish conformal graviton and conformal vector field, we are going to obtain spin-1 and spin-2 conformal fields as dynamical fields. For the $sl(5)$ algebra in our decomposition, we are going to get conformal spin-2 and spin-3 fields, i.e. we are going to get a conformal HS theory.

\section{Holography}

Here, I am going to talk about holography of these theories \cite{Lovrekovic:2023xsj} which we are going to compare. 

The principle of holography says that  we can approach the boundary from the theory in the bulk, and obtain the quantities in the limit which is at that boundary, while at the boundary there is a quantum field theory which will give the same results for the corresponding quantities. 
First we have to split the action in the 2+1 decomposition $\omega=\omega_t dt+\omega_i dx^i$. The action then reads
\begin{align}
I=\frac{k}{4\pi}\int_{\mathcal{R}}dt\int_{\Sigma}d^2x\epsilon^{ij}(g_{ab}\dot{\omega}_i^a\omega_j^b+\omega_t^a F_{ij}^b)+B(\partial\Sigma).
\end{align}
Here, $\Sigma$ denotes two dimensional spatial manifold with $2N$ dynamical fields $\omega_i$, for $N$ dimension of the gauge group $G$, and $N$ Lagrange multipliers $\omega_0$. Then, we have to vary the action to verify that it has a well defined variational principle.  The field equations of the Lagrange multipliers lead to first class constraints and generate gauge transformations. The well known procedure \cite{Campoleoni:2010zq, Blagojevic:2002du, Henneaux:2010xg}, then leads to the charge which tells us about the properties of the theory at the boundary. For the gauge parameter $\Lambda$ independent of the fields, the boundary term has the form 
\begin{align}
Q(\Lambda)=-\frac{k}{2\pi}\int_{\partial\Sigma}dx^i \text{tr}(\Lambda \omega_i)\label{charge}
\end{align}
and after gauge fixing and solving the constraints it defines the global charges of the CS theory.
We partially fix the gauge by setting $\delta \omega_i^{\rho}=0$ and fixing the projector $b(\rho)=\exp^{f(\rho)}$. The invariance of the gauge field $\omega$ is now taken into account while $\delta\omega_i^{\rho}$ and $b(\rho)$ are partially fixing the gauge, 
\begin{align}
\omega(\rho,t,\phi)=b(\rho)^{-1}(\text{d}+\Omega(t,\phi))b(\rho).
\label{gaugecond}\end{align}
We impose near horizon boundary conditions (b.c.s) by expanding $\Omega(t,\phi)$ in generators that are diagonal in the representation that we are considering. 

On the example of $so(3,2)$ we write the b.c.s as
\begin{align}
\Omega(t,\phi)=(\mathcal{K}(t,\phi)d\phi+\mu(t,\phi)dt)D+(\mathcal{K}_d(t,\phi)+\mu_d(t,\phi))L_d,\label{bcs1}
\end{align}
here $\mathcal{K}$ and $\mak_d$ are dynamical fields, $D$ and $L_d$ are diagonal generators, and $\mu$ and $\mu_d$ are chemical potentials. Chemical potentials are fixed at the boundary $\delta\mu=0$. The gauge parameters are expanded in the analogous way $\Lambda=\lambda D+\lambda_d L_d$.
To fix them, as mentioned above, we solve the equations of motion, and gauge transformation conditions.
Using the equation (\ref{charge}) we obtain charges which correspond to integrals over the functions $\mathcal{K}$, $\mathcal{K}_d$, and corresponding gauge parameters
\begin{align}
Q[\lambda,\lambda_d]=\frac{k}{2\pi}\oint d\phi(2\lambda(\phi)\mathcal{K}(\phi)+\frac{3}{2}\lambda_d(\phi)\mathcal{K}_d(\phi)).
\end{align}
The canonical generators fulfill $\delta_{\Lambda_1}Q[\Lambda_2]=\{Q[\Lambda_2],Q[\Lambda_1]\}$, and the charges form an asymptotic symmetry algebra. 
The asymptotic symmetry algebra that is obtained for the near horizon b.c.s is the algebra of $u(1)$ currents at the boundary. In a physical sense, they are interpreted as a hair near the black hole horizon \cite{Hawking:2016msc,Afshar:2016wfy}.

\subsection{First few theories from the tower of theories in new conformal HiSGRA in 3d}
The first theory in the tower of new conformal HiSGRA theories in 3d, is conformal gravity, i.e. spin-2 theory. The following theory we obtain if we add the algebra of conformal vector to conformal algebra.  Conformal algebra and algebra of conformal vector are isomorphic to $sl(4)$, therefore, we can impose the same b.c.s to $sl(4)$ gauge theory and calculate the boundary charges and asymptotic symmetry algebra. The result is going to be analogous to the result for spin-2 theory, however it is going to have an observable for one extra gauge field, because in fundamental representation of $sl(4)$ algebra in 3d, we have three generators represented with a diagonal matrix. This is where we see the importance of the embedding of $so(3,2)$ in $sl(4)$ if we want to relate the lower and higher spin theories from this tower of theories.

The theory with conformal spin-3 field and a conformal spin-2 field is isomorphic to $sl(5)$. It is going to have results for one further gauge field. In this case we will obtain four charges at the boundary, and the asymptotic symmetry algebra that they define. 

Another observable that we can consider is the entropy. Our construction and boundary conditions led to black hole solution. A full clarification of higher spin black hole thermodynamics was
given in \cite{Bunster:2014mua}.  
  The entropy of the higher spin black holes is  \cite{Grumiller:2016kcp, Perez:2013xi, Bunster:2014mua}  
\begin{align}
    S=-\frac{k_N}{\pi}Im\left(\beta\int d\phi Tr(\omega_{\phi}\omega_{\tau})\right).\label{entropygen}
\end{align} 
Here, $\beta$ is the inverse temperature, $k_N$ is a constant, and $\tau=it$ for $0\leq\tau<\beta$. This will give us the expression in terms of the fields and chemical potentials. The values of chemical potentials we can determine by calculating the holonomy around a contractible circle. 
For the first three theories in the tower of new conformal HiSGRA in 3d these are given in Table 1.

\begin{table}[ht!]
\begin{center}
\begin{tabular}{|c|l|l|}
\hline
Gauge algebra &  Dynamical fields  & Entropy   \\ \hline
$so(3,2)$ & $\mathcal{K}, \mathcal{K}_d$  & $\begin{matrix}S=-2k_N \left[ -(m+n+1)\pi\mathcal{K} \right.  \\ -\left.(m-n)\pi\mathcal{K}_d \right]\end{matrix}$  \\ \hline
$sl(4)$ & $\tilde{\mathcal{K}}, \mathcal{K}_0^{(3)}, \mathcal{K}_0^{(4)}$ & $\begin{matrix}S=-\frac{4}{5}\pi k_N \left(5 (l-m-2 n-  1) \tilde{\mak}\right. \\+\left.6  (2 l+ 3 m+n+3)\mak_0^{(4)}\right.\\ \left.+20  (l+n+1 )\mak_0^{(3)}\right)\end{matrix}$\\ \hline 
$sl(5)$ &$\tilde{\mathcal{K}}, \mathcal{K}_0^{(3)}, \mathcal{K}_0^{(4)},\mathcal{K}_0^{(5)}$ & $\begin{matrix} S=-2k_N\pi\bigl[2(p-k-2m-3n)\tilde{\mak}\\ +4\left(3(n+p)-k\right)\mak_0^{(3)} \\ +\frac{24}{5}(2k+4m+n+3p)\mak_0^{(4)}\nonumber \bigr.\bigl.\\+\frac{48}{7}(2k+n+p)\mak_0^{(5)} \bigr]\end{matrix}$
 \\ \hline
\end{tabular}
\caption{Gauge algebra, dynamical fields that appear by imposing near horizon b.c.s, and entropy, for the first three theories in the tower of new conformal HiSGRA in 3d. Here, $m,n\in \mathbb{N}$.}
\end{center}
\end{table}
The fields $\tilde{\mak}$, and $\mak_{0}^{(a)}$, for $a=3,4,5$, are in fundamental representation, and they are linear combinations of the fields in the representation that distinguishes conformal graviton field from conformal spin-s fields, which we usually call PKLD representation.
From the table, we can notice that by switching off the fields that are appearing due to additional generators in the algebra, we will reduce the entropy to be proportional to the entropy of the BTZ black hole (BH). This is visible after taking into account the used embedding.

\section{Metric formulation}

In order to transition to the metric formulation we need to determine the group element $b(\rho)$ and read out the components of the connection $\omega_{\mu}$ which are multiplying the generator of translations in the expansion (\ref{gaugecond}). The general form of the connection is then 
$\omega_{\mu}=e_{\mu}^aP_a+\tilde{\omega}_{\mu}^aJ_a+f_{\mu}^aK_a+b_{\mu}D$. Here $P_a,J_a,K_a,D$ are generators of translations, Lorentz rotations, special conformal transformations, and dilatations, respectively. They are multiplied with corresponding dynamical field. The field $e^a_{\mu}$ is a dreibein and 
the components of dreibein define the metric via $g_{\mu\nu}=e^a_{\mu}e^b_{\nu}\eta_{ab}$. For $\eta_{ab}$ a flat metric. 

From (\ref{bcs1}) we can see that the initial set of b.c.s does not contain the generator of translations. In order to have the generator of translations in the final form of $\omega_{\mu}$, the group element $b(\rho)$
that we choose, will have to contain generator which commutes into $P_{a}$ with one of the generators in boundary conditions (\ref{bcs1}).
In that case, after inserting $b(\rho)$ in (\ref{gaugecond}), we will obtain an invertible $e_{\mu}^a$.

We choose the group element to be 
\begin{align}
b(\rho)=e^{Y(\rho)}, && Y(\rho)=a_1(\rho)D+a_2(\rho)P_y+a_3(\rho)P_x,
\end{align}
 write (\ref{gaugecond}) by using the Baker-Campbell-Hausdorff formula, and calculate $g_{\mu\nu}$.

The general form of the metric that we obtain, we can bring into more familiar form. By specifying functions $a_1(\rho),a_2(\rho)$ and $a_3(\rho)$, and transforming the coordinates, we can write the metric as

 \begin{align}
 ds^2=f_{\rho\rho}(\rho)d\rho^2+f_{tt}(\rho)dt^2-2\rho^2n(\rho)dtd\phi+\rho^2d\phi^2\label{solution}
 \end{align}
 for 
\besubeqs \label{metricso32}
 \begin{align}
 f_{\rho\rho}(\rho)&=\frac{\rho^2}{(\mak \rho c_1-\mak-\rho)(\mak \rho c_1 -\mak+\rho)(\mak_d\rho c_1-\mak_d-\rho)(\mak_d \rho c_1-\mak_d+\rho)} \label{horizons}\\
 n(\rho)&=\frac{\mak\mu-\mak_d\mu_d}{\mak\mu_d-\mak_d\mu}-\frac{\rho(\mak\mak_dc_1(\rho c_1-2)-\rho c_2)+\mak\mak_d}{\rho^2}\\
 f_{tt}&=\rho^2 n(\rho)^2-\frac{1}{f_{\rho\rho}(\rho)}.
 \end{align}
\esubeqs
Here, $c_1$ and $c_2$ are constants. 
This is not the most general solution this procedure gives for the group element (3.1). The most general solution contains a function with an arbitrary dependence on $\rho$. However,  (3.2), is an interesting solution on which we can demonstrate special cases from the literature. 
The Ricci scalar of the solution (\ref{solution}) is
\begin{footnotesize}
\begin{align}
R=-6+\frac{2 c_1}{\rho^3}\left[\left( -2\mak^2(\mak_d^2+\rho^2) \right)+\rho c_1 \left(3\rho^2(\mak^2+\mak_d^2)+\mak^2\mak_d^2 \rho c_1(4-3 \rho c_1)+\mak^2\mak_d^2 \right)-2\mak_d^2\rho^2\right].
\end{align}
\end{footnotesize}
We can see that it can be reduced to AdS solution when the constant $c_1$ is set to zero. This, together with taking $c_2=0$ will correspond to the situation 3. listed below. 

We can identify the following solutions
\begin{enumerate}
\item from the (\ref{horizons}) we can see that for $c_1>\frac{1}{\mak}$, $c_1>\frac{1}{\mak_d}$ and $\mak,\mak_d>0$, we have four positive eigenvalues. 
\item If we set $\mak_0, c_2=\frac{\mu_d}{\mu}$ we will get rescaled 3d analog of Mannheim-Kazanas-Riegert (MKR) solution \cite{Mannheim:1988dj,Riegert:1984zz}, which is Oliva-Tempo-Troncoso (OTT) solution \cite{Oliva:2009hz}. The subcases of the OTT solution include; 
 locally flat BH;  dS BH;  AdS BH; and  extremal BH. 
\item If we choose the $\mak$ and $\mak_d$ to be the same, and 1. to hold, we get two eigenvalues, similarly as if when we  choose $c_1=c_2=0$.
\item When we set $a_1(\rho)=a_2(\rho)=1$ in the dreibein, and $a_3(\rho)=c_2+c_1\rho$ we can bring the metric in the Lobachevsky form \cite{Bertin:2012qw}. In the Lobachevsky metric we can bring the solution to the form of 
\begin{itemize}
\item global Lobachevsky: for $c_1=1,\mak_d=0$
\item Poincar\'e Lobachevsky: for $c_1=1,\mak_d=\mak=0$
\item rotating Lobachevsky: for $c_1=1,\mak_d=\mak$.
\end{itemize}
\end{enumerate}

In the same way as we determined the solution (\ref{solution}), we can find the solution for the gauge theory for $sl(4)$ group. The representation for $sl(4)$ group that we use is obtained by adding a set of ten matrices for $so(3,2)$ and a set of five matrices that are a representation of conformal vector. This means that the solution (\ref{solution}) will be generalized with the charges coming from conformal vector. In the solution that will manifest so that in addition to $\mak$, $\mak_d$ we get additional $\mak_c$ coming from the conformal vector. 

However, since now we have an entire $sl(4)$ we can express the solution in terms of $\tilde{\mak}$, $\mak_{0}^{(3)}$ and $\mak_{0}^{(4)}$. To do this we need to know the map between the PKLD representation of conformal graviton with conformal vector, and fundamental representation of $sl(4)$.

Both of these representations have its advantages and disadvantages. Since $so(3,2)$ algebra is not isomorphic to one of the $sl(N)$ algebras, the advantage of the representation with conformal graviton and conformal vector field is the manifest distinction between the $so(3,2)$ gauge theory and the theory with $so(3,2)$ and conformal vector field, as well as $so(3,2)$ and conformal spin-3 field, etc. 
On the other side the representation that uses $sl(N)$ matrices allows for comparison with results for $sl(N)$ theories, most studied in the literature. 


When we use the consistent embedding of the $so(3,2)$ in the $sl(N)$ algebras, we can classify our solutions with respect to one parameter subgroups of $so(3,2)$.
That includes the classification with respect to the solution to which the higher spin solution reduces when the additional charges are set to zero.

\section{Classification of solutions according to one parameter subgroups of $so(3,2)$}

The tensors in $so(3,2)$ are described by antisymmetric matrices $t_{ab}=-t_{ba}$. We can determine the types of these matrices based on their eigenvalues. If we denote the generators of $so(3,2)$ group with \begin{align}
    J_{ab}=x_b\frac{\partial}{\partial x_a}-x_a\frac{\partial}{\partial x_b} 
\end{align}
 for the coordinates $x^a=(u,v,x,y,z)$, we can write the most general Killing vector as  \\ $\xi=\frac{1}{2}t^{ab}J_{ab}$. 

 These Killing vectors generate different types of solutions between which we cannot transform using $so(3,2)$ transformations.  The Casimir invariants of the solutions remain the same after the transformation, which allows us to determine to which type the transformed solution belongs.
If we find the transformation of coordinates from one form of the solution to another, the distinctive features that tell us the class of the solution, are therefore Casimir invariants.

In Table 2. we can see the classification according to one parameter subgroups of $so(3,2)$.

\begin{table}[ht!]
    \centering
    \begin{tabular}{|c|c|c|c|}
\hline         Type& Killing vector & 1. Casimir & 3. Casimir \\ \hline
       $  I_a$&$b(J_{01}+J_{23})-a(J_{03}+J_{12})$& $4(b^2-a^2)$ & $4(a^3-3ab^2)$\\
         $I_b$ & $\lambda_2 J_{03}+\lambda_1 J_{12}$ & $-2(\lambda_1^2+\lambda_2^2)$ & $2(\lambda_1^3+\lambda_2^3) $\\
         $I_c$ & $b_1 J_{01}+b_2 J_{23}$ & $2(b_1^2+b_2^2)$ &0 \\
         $I_d$ & $b J_{01}+\lambda J_{03}$ & $2(b^2-\lambda^2)$ &  $ 2\lambda(b_1^2-\lambda^2)$ \\
  $ II_a$ & $J_{01}-J_{02}-J_{13}+J_{23}-\lambda(J_{03}+ J_{12})$& $ -4\lambda^2$ & $4\lambda^2(3+\lambda)$\\ 
         $II_b$ & $(b-1)J_{01}+(b+1)J_{32}+J_{02}-J_{13}$ & $4b^2$ & 0\\
         $III_{a+}$ & $-J_{13}+J_{23}$   & 0& 0 \\
         $III_{a-}$ & $-J_{01}+J_{02}$ &0 & 0\\
         V & $\frac{1}{4}(-J_{01}+J_{03}-J_{12}-J_{23})+J_{04}+J_{24}$ & 0& -2\\ \hline
         \end{tabular}
    \caption{One-parameter subgroups of $so(3,2)$}
    \label{tabl1}
\end{table}

$a$, $b$, $b_1$, $b_2$, $\lambda_1$, and $\lambda_2$ are real numbers determined by the eigenvalues of $t^{ab}$, which form the basis for the classification. Specifically, $\lambda_1$ and $\lambda_2$ represent real eigenvalues, whereas $a$, $b$, $b_1$, and $b_2$ originate from the complex eigenvalues.

In comparison to the solutions of the Einstein gravity  in 3d \cite{Banados:1992gq}, the solutions of the conformal gravity in 3d have two additional types in the classification. These are types $I_d$ and $V$. Type $I_d$  has one real and one imaginary eigenvalue. The example of the solution that belongs to the $I_d$ type was given in \cite{Lovrekovic:2023xsj}. Type $V$ indicates that it might describe some special extremal solution.

\section{Summary}

In this talk we have seen that taking into account conformal symmetry in addition to standard Poincar\'e symmetries of Einstein gravity, in three dimensions, leads to more general solutions in comparison to BTZ. 
Concretely, we have talked about a solution which contains four horizons. 

We have explained that with a specific choice of a group projector, and specific parameters we can restrict the general solution to: (i) solution with functional dependency on $\rho$, (ii) solution with four horizons, (iii) BTZ solution, (iv) OTT solutions, and (v) Lifshitz solutions. Solutions of conformal gravity can be classified according to one parameter subgroups of $so(3,2)$ and they can realise two new classes which do not appear in the classification of the one parameter subgroups of $so(2,2)$. These classes are $I_d$ and $V$ from the Table 2.

Depending on the embedding of the $so(3,2)$ algebra into $sl(4)$, $sl(5)$ and further algebras from the tower of algebras that define new conformal HiSGRA in 3d, we can generalize the above solutions to higher spin solutions. 

When in such solutions we switch off the higher spin fields, we should arrive to the conformal gravity solutions. In this way, we can classify the HS solutions based on the conformal gravity solutions they connect to.



\bibliographystyle{JHEP-2}
\bibliography{megabib.bib}

\providecommand{\href}[2]{#2}\begingroup\raggedright\begin{thebibliography}{10}

\bibitem{Chern:1974ft}
S.-S. Chern and J.~Simons, {\it {Characteristic forms and geometric
  invariants}},  {\em Annals Math.} {\bf 99} (1974) 48--69.

\bibitem{Dunne:1998qy}
G.~V. Dunne, {\it {Aspects of Chern-Simons theory}},  in {\em {Les Houches
  Summer School in Theoretical Physics, Session 69: Topological Aspects of
  Low-dimensional Systems}}, 7, 1998.
\newblock \href{http://arXiv.org/abs/hep-th/9902115}{{\tt hep-th/9902115}}.

\bibitem{Witten:1988hc}
E.~Witten, {\it {(2+1)-Dimensional Gravity as an Exactly Soluble System}},
  {\em Nucl. Phys.} {\bf B311} (1988) 46.

\bibitem{Grigoriev:2019xmp}
M.~Grigoriev, I.~Lovrekovic and E.~Skvortsov, {\it {New Conformal Higher Spin
  Gravities in $3d$}},  {\em JHEP} {\bf 01} (2020) 059
  [\href{http://arXiv.org/abs/1909.13305}{{\tt 1909.13305}}].

\bibitem{Lovrekovic:2023xsj}
I.~Lovrekovic, {\it {Holography of New Conformal Higher Spin Gravities in 3d}},
   \href{http://arXiv.org/abs/2312.12301}{{\tt 2312.12301}}.

\bibitem{Campoleoni:2010zq}
A.~Campoleoni, S.~Fredenhagen, S.~Pfenninger and S.~Theisen, {\it {Asymptotic
  symmetries of three-dimensional gravity coupled to higher-spin fields}},
  {\em JHEP} {\bf 1011} (2010) 007 [\href{http://arXiv.org/abs/1008.4744}{{\tt
  1008.4744}}].

\bibitem{Blagojevic:2002du}
M.~Blagojevic, {\em {Gravitation and gauge symmetries}}.
\newblock 2002.

\bibitem{Henneaux:2010xg}
M.~Henneaux and S.-J. Rey, {\it {Nonlinear $W_{infinity}$ as Asymptotic
  Symmetry of Three-Dimensional Higher Spin Anti-de Sitter Gravity}},  {\em
  JHEP} {\bf 1012} (2010) 007 [\href{http://arXiv.org/abs/1008.4579}{{\tt
  1008.4579}}].

\bibitem{Hawking:2016msc}
S.~W. Hawking, M.~J. Perry and A.~Strominger, {\it {Soft Hair on Black Holes}},
   {\em Phys. Rev. Lett.} {\bf 116} (2016), no.~23 231301
  [\href{http://arXiv.org/abs/1601.00921}{{\tt 1601.00921}}].

\bibitem{Afshar:2016wfy}
H.~Afshar, S.~Detournay, D.~Grumiller, W.~Merbis, A.~Perez, D.~Tempo and
  R.~Troncoso, {\it {Soft Heisenberg hair on black holes in three dimensions}},
   {\em Phys. Rev. D} {\bf 93} (2016), no.~10 101503
  [\href{http://arXiv.org/abs/1603.04824}{{\tt 1603.04824}}].

\bibitem{Bunster:2014mua}
C.~Bunster, M.~Henneaux, A.~Perez, D.~Tempo and R.~Troncoso, {\it {Generalized
  Black Holes in Three-dimensional Spacetime}},  {\em JHEP} {\bf 05} (2014) 031
  [\href{http://arXiv.org/abs/1404.3305}{{\tt 1404.3305}}].

\bibitem{Grumiller:2016kcp}
D.~Grumiller, A.~Perez, S.~Prohazka, D.~Tempo and R.~Troncoso, {\it {Higher
  Spin Black Holes with Soft Hair}},  {\em JHEP} {\bf 10} (2016) 119
  [\href{http://arXiv.org/abs/1607.05360}{{\tt 1607.05360}}].

\bibitem{Perez:2013xi}
A.~Perez, D.~Tempo and R.~Troncoso, {\it {Higher spin black hole entropy in
  three dimensions}},  {\em JHEP} {\bf 04} (2013) 143
  [\href{http://arXiv.org/abs/1301.0847}{{\tt 1301.0847}}].

\bibitem{Mannheim:1988dj}
P.~D. Mannheim and D.~Kazanas, {\it {Exact Vacuum Solution to Conformal Weyl
  Gravity and Galactic Rotation Curves}},  {\em Astrophys. J.} {\bf 342} (1989)
  635--638.

\bibitem{Riegert:1984zz}
R.~J. Riegert, {\it {Birkhoff's Theorem in Conformal Gravity}},  {\em Phys.
  Rev. Lett.} {\bf 53} (1984) 315--318.

\bibitem{Oliva:2009hz}
J.~Oliva, D.~Tempo and R.~Troncoso, {\it {Static spherically symmetric
  solutions for conformal gravity in three dimensions}},  {\em Int. J. Mod.
  Phys. A} {\bf 24} (2009) 1588--1592
  [\href{http://arXiv.org/abs/0905.1510}{{\tt 0905.1510}}].

\bibitem{Bertin:2012qw}
M.~Bertin, S.~Ertl, H.~Ghorbani, D.~Grumiller, N.~Johansson and D.~Vassilevich,
  {\it {Lobachevsky holography in conformal Chern-Simons gravity}},  {\em JHEP}
  {\bf 06} (2013) 015 [\href{http://arXiv.org/abs/1212.3335}{{\tt 1212.3335}}].

\bibitem{Banados:1992gq}
M.~Banados, M.~Henneaux, C.~Teitelboim and J.~Zanelli, {\it {Geometry of the
  (2+1) black hole}},  {\em Phys. Rev. D} {\bf 48} (1993) 1506--1525
  [\href{http://arXiv.org/abs/gr-qc/9302012}{{\tt gr-qc/9302012}}]. [Erratum:
  Phys.Rev.D 88, 069902 (2013)].

\end{thebibliography}\endgroup

\end{document}